\documentclass[10pt,preprint]{sigplanconf}

\usepackage[utf8]{inputenc}

\usepackage[english]{babel}

\usepackage{microtype}

\usepackage[pdftex,final]{graphicx}

\usepackage{listings}
\usepackage{float}
\usepackage[usenames,dvipsnames]{xcolor}

\usepackage[nolist]{acronym}

\usepackage{arydshln}

\usepackage{tikz}
\usetikzlibrary{shapes}
\newcommand\textcircle[1]{%
  \tikz[baseline=(X.base)]
    \node (X) [draw, shape=circle, inner sep=-0.5] {\strut #1};}
\newcommand\textdiamond[1]{%
  \tikz[baseline=(X.base)]
    \node (X) [draw, shape=diamond, inner sep=-0.5] {\strut #1};}

\usepackage[%
            pdfauthor={Tim Besard},%
            pdftitle={High-level GPU programming in Julia},%
            pdfkeywords={Julia, CUDA}]%
            {hyperref}
\usepackage[all]{hypcap}    %for going to the top of an image when a figure reference is clicked

\usepackage{cleveref}

\begin{document}

%
% Language definitions
%

% From ColorBrewer
\definecolor{lst1}{RGB}{228,26,28}
\definecolor{lst2}{RGB}{55,126,184}
\definecolor{lst3}{RGB}{77,175,74}

\lstset{%
  basicstyle          = \ttfamily\footnotesize,
  keywordstyle        = \color{lst2}\underline,
  stringstyle         = \color{lst1},
  commentstyle        = \itshape\color{lst3},
  showstringspaces    = false,
  frame               = top,
  frame               = bottom,
  framextopmargin     = 2pt,
  framexbottommargin  = 2pt,
  framexleftmargin    = 17pt,
  xleftmargin         = 17pt,
  belowskip           = 0ex,
  numbers             = left,
  numbersep           = 7pt,
  escapechar          = \&
}

\lstdefinelanguage{Julia}{%
  morekeywords={%
    abstract,begin,break,case,catch,const,continue,do,else,elseif,end,export,%
    false,for,function,immutable,import,importall,if,in,macro,module,%
    otherwise,quote,return,switch,true,try,type,typealias,using,while},%
  sensitive=true,%
  alsoother={\$},%
  morecomment=[l]\#,%
  morecomment=[n]{\#=}{=\#},%
  morestring=[s]{"}{"},%
  morestring=[m]{'}{'},%
}[keywords,comments,strings]

% NOTE: normally we'd use `morekeywords`, but this time we only want to
%       highlight the keywords specific to CUDA or CUDA.jl

\lstdefinelanguage{CUDA}[ANSI]{C}{%
  keywords={%
    __global__,__device__,%
    threadIdx,blockIdx},%
}

\lstdefinelanguage{CUDA.jl}[]{Julia}{%
  keywords={%
    CuDevice,CuContext,CuModule,CuFunction,CuArray,%
    launch,to_host,free,unload,destroy,create_context,%
    @target,@cuda,blockId_x,threadId_x,numBlocks_x},%
}

%
% Acronyms
%

\begin{acronym}

\acro{ir}[IR]{Intermediate Representation}

\acro{ptx}[PTX]{Parallel Thread Execution}
\acro{ast}[AST]{Abstract Syntax Tree}

\acro{jit}[JIT]{Just-in-Time}
\acro{isa}[ISA]{Instruction Set Architecture}

\end{acronym}

%
% Paper properties
%

\setlength{\pdfpageheight}{\paperheight}
\setlength{\pdfpagewidth}{\paperwidth}

\conferenceinfo{CGO '16}{March 12--18, 2016, Barcelona, Spain}
\copyrightyear{2016}
\copyrightdata{978-1-nnnn-nnnn-n/yy/mm}
\doi{nnnnnnn.nnnnnnn}

% Uncomment one of the following two, if you are not going for the
% traditional copyright transfer agreement.

%\exclusivelicense                % ACM gets exclusive license to publish,
                                  % you retain copyright

%\permissiontopublish             % ACM gets nonexclusive license to publish
                                  % (paid open-access papers,
                                  % short abstracts)

% Preprint specific text
\titlebanner{}
\preprintfooter{}

\title{High-level GPU programming in Julia}
%\subtitle{Subtitle Text, if any}

\authorinfo{Tim Besard}
           {Computer Systems Lab\\Ghent University, Belgium}
           {Tim.Besard@elis.ugent.be}
\authorinfo{Pieter Verstraete}
           {Ghent University, Belgium}
           {}
\authorinfo{Bjorn De Sutter}
           {Computer Systems Lab\\Ghent University, Belgium}
           {Bjorn.DeSutter@elis.ugent.be}

\maketitle

%%%%%%%%%%%%%%%%%%%%%%%%%%%%%%%%%%%%%%%%%%%%%%%%%%%%%%%%%%%%%%%%%%%%%%%%%%%%%%%%
% Contents
%

\begin{abstract}

%% Foreword (the before)
%
% context (why the need is pressing or important)
GPUs are popular devices for accelerating scientific calculations.
%
% need (why something needed to be done)
However, as GPU code is usually written in low-level languages, it breaks the
abstractions of high-level languages popular with scientific programmers.
%
% task (what was undertaken to address the need) + object (what the present
% document does or covers)
To overcome this, we present a framework for CUDA GPU programming in the
high-level Julia programming language. This framework compiles Julia source code
for GPU execution, and takes care of the necessary low-level interactions using
modern code generation techniques to avoid run-time overhead.

%% Summary (the after)
%
% findings (what the work yielded)
Evaluating the framework and its APIs on a case study comprising the trace
transform from the field of image processing, we find that the impact on
performance is minimal, while greatly increasing programmer productivity. The
metaprogramming capabilities of the Julia language proved invaluable for
enabling this.
%
% conclusions (what the findings mean)
Our framework significantly improves usability of GPUs, making them accessible
for a wide range of programmers.
It is available as free and open-source software licensed under the MIT License.

\end{abstract}

% Overview of categories: http://www.acm.org/about/class/ccs98-html


% general terms are not compulsory anymore,
% you may leave them out
%\terms
%term1, term2


\section{Introduction}
\label{introduction}
%
%

% ex-TODO: it would also have been interesting to compare against typical 2-tier
% run-time API's and the accompanying overhead, either semantically (i.e.
% combinatorial explosion in terms of which method to call, e.g. cuFFT
% `cufftExecC2C`, `cufftExecZ2Z`, `cufftExecR2C`, etc) or performance (e.g.
% BLAS, LAPACK: first argument indicates the type).

GPUs can significantly speed up certain workloads. However, targeting GPUs
requires serious effort. Specialized machine code needs to be generated
through the use of a vendor-supplied compiler. Because of the architectural
set-up, initiating execution on the coprocessor is often quite complex as well.
Even though the vendors try hard to supply toolchains that support different
developer environments and offer convenience functionality to lower the burden,
they are essentially playing catch-up.

While coprocessor hardware improves program efficiency, high-level languages are
becoming a popular choice because of their improved programmer productivity.
Languages such as Python or Julia provide a user-friendly development
environment. Low-level details are hidden from view, and secondary tasks such as
dependency management and compiling and linking are automatically taken care of.

For users of these high-level languages, jumping through the many hoops of GPU
development is often an exceptionally large burden. A lot of low-level knowledge
is required, and many of the user-friendly abstractions break down. For example,
when using Python to target NVIDIA GPUs using the CUDA toolkit, the developer
needs to write GPU kernels in CUDA~C, and interact with the CUDA API in order to
compile the code, prepare the hardware and launch the kernel. The situation is
even worse for languages unsupported by the CUDA toolkit, such as Julia, in
which case there are only superficial or no CUDA API wrappers at all.

% Requirements

Ideally, it should be possible to develop and execute high-level GPU kernels
without much extra effort: writing kernels in high-level source code, while the
interpreter for that language takes care of compiling the necessary functions to
GPU machine code. Low-level details should be automated, or at least wrapped in
user-friendly language constructs.

% Contributions

This paper presents a framework to target NVIDIA GPUs, and by extent other
accelerators, directly in the Julia programming language: Kernels can be written
in high-level Julia code. We also created high-level CUDA API wrappers to
support the natural use of the CUDA API from within Julia. The framework
provides a user-friendly GPU kernel programming and execution interface that
automates driver interactions and abstracts GPU-specific details without
introducing any run-time overhead. All code implementing this framework is
available as open-source code on GitHub.

% Structure

In \Cref{background} we describe relevant technologies and the motivation for
our work. \Cref{overview} provides an overview of our framework, each component
explained in detail in \Cref{julia-kernels,api-wrapper,automation}. Finally, we
evaluate our work in \Cref{evaluation}.

\section{Motivation and Background}
\label{background}

\subsection{CUDA programming}
\label{background:cuda}

Executing code on a coprocessor like a GPU requires developers to send specific
machine code over to the GPU, prepare the execution environment (configure
hardware properties, upload required data to device memory, etc.) and finally
start execution. All these operations are performed by calling into the device
driver, creating the explicit distinction between \emph{host code} running on
the CPU, responsible for configuring the environment, and \emph{device code}
running on the GPU, performing the actual computations.

The hardware vendor often provides a user-friendly toolchain that takes care of
some of these details. In the case of NVIDIA GPUs for example, the CUDA toolkit
allows writing GPU kernels in C or C++, and provides system headers that wrap
the necessary CUDA driver calls in more user-friendly, language-native
constructs. Such toolchains are often only available for systems languages, and
still require significant knowledge about the platform and its specifics.

The CUDA toolkit comes with two APIs: the low-level driver API, and the
higher-level run-time API implemented on top of the driver
API~\cite{nvidia2015cudadoc}. We have implemented our framework using the driver
API, because certain low-level functionality is missing from the run-time API.

Launching a kernel using the driver API consists at least of creating a code
module, extracting a function handle and launching the kernel given this handle
and other parameters like grid and block dimensions, shared memory, kernel
arguments, etc. Code modules are created by sending \ac{ptx} code containing one
or more kernels to the API. \ac{ptx} code is comparable to traditional CPU
assembly code, but it is a virtual \ac{isa}, translated by the device driver to
the target \ac{isa}. This hides device-specific properties to improve
portability.

\subsection{Julia technical computing language}
\label{background:julia}

Julia is a dynamic language for scientific computing, designed for
performance~\cite{bezanson2014julia}. By means of aggressive code specialization
against run-time types in combination with a \ac{jit} compiler using the LLVM
compiler infrastructure~\cite{lattner2004llvm}, the compiler is able to emit
highly efficient machine code. This makes it possible to write
performance-sensitive code in Julia itself, avoiding the classical two-tier
architecture (also dubbed ``the two language
problem''~\cite{bezanson2014julia}), where application logic is expressed in a
high-level language but heavy computations are performed in a low-level
programming language like C or C++. To illustrate this point, the entire Julia
standard library is written in Julia itself (with some obvious exceptions for
the purpose of reusing existing libraries), while still offering good
performance.

\subsubsection{Compiler overview}
\label{background:julia:compiler}

Julia code is not statically compiled. When invoking the Julia interpreter to
execute a source file, the code will seemingly be executed without a separate
compilation step, much like other scripting languages such as Python or R.
However, behind the scenes a \ac{jit} compiler generates native machine code,
employing static code generation techniques to improve performance. In the
current situation, this Julia code generation is limited to LLVM-supported CPUs.

\begin{figure}[!t]
  \centering
  \includegraphics{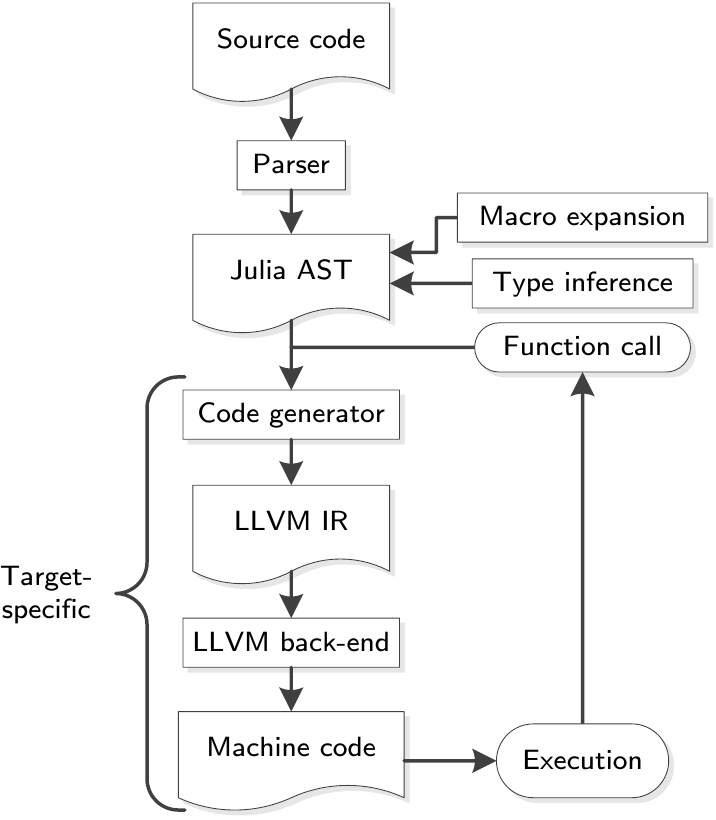}
  \caption{Julia compiler components and interactions.}
  \label{fig:compiler}
\end{figure}

\Cref{fig:compiler} shows the main components for compiling and executing Julia
source code. First, source code is parsed into an \ac{ast}~\cite{julia2015doc}.
Then, metaprogramming constructs are expanded (see
\Cref{background:julia:metaprogramming}), and type inference takes place,
annotating the \ac{ast} for optimization and code generation purposes. The typed
\ac{ast} is used by the code generator to emit LLVM \ac{ir} code, which finally
can be compiled to machine code using one of the available LLVM back-ends.

Note that the resulting execution can trigger another compilation. This happens,
e.g., when executing a call to a function that was not defined at the point the
call was generated. The called function will then be compiled at run-time.
On-the-fly compilation is also used to type-specialize functions, triggering
recompilation when the arguments' types change.

Whereas the Julia \ac{ast} is target-independent, the code generator contains
conditional code paths depending on the type of processor. These conditionals
take care of platform-dependent functionality, and cope with the fact that LLVM
\ac{ir} is not target independent (due to explicit pointer widths, memory
alignment, calling conventions, etc.).

\subsubsection{Two-tier GPU support}
\label{background:julia:gpu}

Although the Julia toolchain cannot generate code for a GPU, it is possible to
target such devices in a classical two-tier approach: write GPU code in a
supported language, and use the resulting compiled objects from within Julia.

{
  \nocaptionrule
  \begin{lstlisting}[float, language=cuda, label={lst:cuda},
                   caption={CUDA~C device code for vector addition.}]
extern "C"
{
  // define a kernel
  __global__ void vadd(const float *a,&\newline&                       const float *b,&\newline&                       float *c)
  {
    int i = threadIdx.x +&\newline&            blockIdx.x * blockDim.x;
    c[i] = a[i] + b[i];
  }
}
  \end{lstlisting}
}

In the case of an NVIDIA GPU, this means writing the code in CUDA~C, compiling
it to \ac{ptx} instructions and using the CUDA APIs to prepare and execute the
kernel. As a running example, we consider the simple task of adding two
floating-point vectors. \Cref{lst:cuda} contains the CUDA~C source code, which
can be compiled with the CUDA toolkit's \texttt{nvcc} compiler.

Before executing the resulting \ac{ptx} object code, the user needs to use the
CUDA API to configure the hardware and prepare execution. There exist several
unofficial CUDA API wrappers for Julia that target both the CUDA driver and
run-time APIs~\cite{julia2015cuda, julia2015cudart}, including the
\texttt{CUDA.jl} package, the starting point of this work. The wrappers make
heavy use of the \texttt{ccall} feature of Julia, which allows calling into C
code with relative ease. It is noteworthy that these \texttt{ccall} expressions
are compiled to the same instructions a native C call would be, so the resulting
overhead is the same as calling a library function from C
code~\cite{julia2015doc}.

{
  \nocaptionrule
  \begin{lstlisting}[float, language=CUDA.jl, label={lst:julia_original},
                     caption={Julia host code for launching a CUDA kernel using
                              the pre-existing \texttt{CUDA.jl} API wrapper.
                              \vspace{2pt}}]
# set-up
dev = CuDevice(0)
ctx = create_context(dev)

# load kernel
md = CuModule("vadd.ptx")
vadd_fun = CuFunction(md, "vadd")

# create some data
dims = (3, 4)
a = round(rand(Float32, dims) * 100)
b = round(rand(Float32, dims) * 100)

# prepare device memory
ga = CuArray(a)
gb = CuArray(b)
gc = CuArray(Float32, dims)

# execute!
len = prod(dims)
launch(vadd_fun, len, 1,&\newline&       (ga, gb, gc))

# download results
c = to_host(gc)

# verify
@assert a+b == c

# clean-up device memory
free(ga)
free(gb)
free(gc)

# tear-down
unload(md)
destroy(ctx)
  \end{lstlisting}
}

The kernel from \Cref{lst:cuda} can be launched from within Julia using the
existing CUDA driver API wrapper~\cite{julia2015cuda}. Although this wrapper
exposes the API using familiar Julia mechanics, it is a shallow wrapper in which
each expression narrowly maps to a few API calls. This makes the resulting host
code verbose and non-obvious, as shown in \Cref{lst:julia_original} where
underlined keywords are part of the \texttt{CUDA.jl} package, and requires
detailed knowledge about the underlying driver API. Our framework automates
these interactions, significantly reducing the required host code to that of
\Cref{lst:julia_new}.

\subsubsection{Julia metaprogramming}
\label{background:julia:metaprogramming}

Julia has strong support for Lisp-like metaprogramming. The language is
homoiconic: Code is represented as a datastructure of the language itself. This
empowers much of the metaprogramming functionality, as a program can transform
and generate its own source code~\cite{julia2015doc}.

Code can be generated and evaluated at every point during program execution, and
two interfaces allow access to a program's \ac{ast} before it is evaluated. The
first is a \emph{macro}, which gets expanded when the code is parsed. The body
of the macro is executed during macro expansion, and any argument to the macro
is passed symbolically. For example, consider the following macro definition and
invocation:
{ \small
\begin{verbatim}
   macro my_AST_transform(input_ast)
     output_ast = transform(input_ast)
     return output_ast
   end

   @my_AST_transform foobar(42)
\end{verbatim}
}
The macro's body will be evaluated right after the invocation is parsed. Its
argument is passed symbolically, i.e., as an \ac{ast}, in this case an \ac{ast}
that encodes the Julia expression \texttt{foobar(42)}. The return value
\texttt{output\_ast} is also an \ac{ast}, and gets included in the program's
\ac{ast} replacing the original subtree of the macro invocation and its
arguments, i.e., of the whole expression
\texttt{@my\_AST\_transform~foobar(42)}. Macros are not limited to transforming
\acp{ast}. They can generate and inject arbitrary code where the macro was
originally invoked, a feature our framework heavily relies on.

\emph{Generated function} constitute the second metaprogramming interface. These
are similar to macros, but they are evaluated after type inference. As a result,
the \ac{ast} passed to the generated function now also includes the type of the
expressions. This allows generating specialized code, avoiding the run-time
overhead of switching between different implementations based on the type of a
variable or expression.

Macros and generated functions are useful to extend the language, i.e., to
define new syntax that the compiler would otherwise not support. We will also
use it to improve performance by replacing potentially recurring run-time
overhead with one-time calculations during code generation.

\section{System Overview}
\label{overview}

Adding GPU support to the Julia language involved modifying multiple compiler
components and adding significant pieces of functionality. We confined as much
functionality as possible to the \texttt{CUDA.jl} package, which only consists
of Julia code. The remaining changes are part of the code generator, which is
written in C++.

\paragraph{Julia GPU kernels}

We extended the Julia language by adding the \texttt{@target} macro that will be
discussed in more detail in \Cref{julia-kernels:target}. A developer can use
this optional macro to indicate which target a function should be compiled for.
\Cref{lst:julia_new:target} in \Cref{lst:julia_new} illustrates its use for a
Julia GPU kernel.

In combination with the \texttt{@target} macro, we added \ac{ptx} support to the
Julia compiler as detailed in \Cref{julia-kernels:codegen}. This allows
compiling Julia kernel functions to \ac{ptx} code, which can be then executed by
means of the CUDA APIs. Finally, we added intrinsic functions that expose this
compiler functionality. This enables any Julia host code to call into the
compiler and generate \ac{ptx} instructions.

\paragraph{CUDA API wrapper}

Starting from the \texttt{CUDA.jl} Julia package~\cite{julia2015cuda}, we
implemented comprehensive support for launching kernels and all related tasks.
We improved the wrapper's compatibility, making it possible to target non-CUDA
hardware with the GPU~Ocelot emulator~\cite{farooqui2011gpuocelot}. These
contributions are described in \Cref{api-wrapper}. Although we focused on
wrapping the complete API, we also extended the available high-level wrappers to
cover one or more API calls with more idiomatic Julia types or expressions. This
improves user-friendliness, for when a developer needs to use the API manually
rather than rely on the automation infrastructure presented in
\Cref{automation}.

\paragraph{High-level automation}

Improving productivity even further, we provide a way to generate API calls
rather than requiring the developer to call the API manually. Using Julia's
strong metaprogramming capabilities, we inspect how a kernel is invoked, and
generate specialized \ac{ptx} code together with all necessary API calls as soon
as the required type information is known. \Cref{lst:julia_new} shows how this
makes the process of calling GPU kernels almost fully transparent, without
adding any run-time overhead.

{
  \nocaptionrule
  \begin{lstlisting}[float, language=CUDA.jl, label={lst:julia_new},
                   caption={Julia source code with high-level GPU kernel
                            and automated driver interactions.
                            \vspace{2pt}}]
# define a kernel
@target ptx function vadd(a, b, c)&\label{lst:julia_new:target}&
  i = blockId_x() +&\newline&      (threadId_x()-1) * numBlocks_x()&\label{lst:julia_new:intrinsics}&
  c[i] = a[i] + b[i]
end

# create some data
dims = (3, 4)
a = round(rand(Float32, dims) * 100)
b = round(rand(Float32, dims) * 100)
c = Array(Float32, dims)

# execute!
len = prod(dims)
@cuda (len, 1)&\label{lst:julia_new:cuda}\newline&      vadd(CuIn(a), CuIn(b), CuOut(c))

# verify
@assert a+b == c
  \end{lstlisting}
}

\section{Julia GPU Kernels}
\label{julia-kernels}

Adding support for writing GPU kernels in high-level Julia code consists of two
components: a modified Julia compiler for generating \ac{ptx} assembly from
Julia source code, and a language extension for developers to indicate which
functions are meant to be compiled that way.

\subsection{\acs{ptx} code generation}
\label{julia-kernels:codegen}

The existing Julia compiler only supports generating code for traditional CPUs.
A single code generator compiles the \ac{ast} to LLVM \ac{ir} code, with some
conditional code paths depending on the type of processor. These conditionals
are configured at compile time, which means that each compiled Julia interpreter
is tied to a specific platform and processor.

Adding support for emitting \ac{ptx} code is quite invasive. Firstly, as the
Julia code generator needs to generate both \ac{ptx} instructions and native
host code, target selection must happen at run time. Because of how LLVM works,
the \ac{ptx} code will have to reside in a separate code module, isolated from
the host module. Secondly, the GPU architecture and execution model differs
significantly from the processors currently supported by Julia. Semantic
differences in the emitted LLVM \ac{ir} can be resolved easily with conditional
code paths in the code generator. For example, LLVM \ac{ir} supports the notion
of \emph{address spaces}, and every pointer indicates which address space its
pointee lives in. In the case of x86, this is a relatively unused feature, and
the Julia code generator always makes use of the default address space~$0$.
\ac{ptx} code by contrast uses these semantics to differentiate between the
different types of device memory, such as global, constant or texture memory.
This required modifying the Julia code generator to make sure that the address
space property is conserved and appropriate glue code is generated when
exchanging values between different address spaces.

A bigger restriction stems from the fact that the GPU and CPU memories are
disjoint, i.e., pointers are not interchangeable\footnotemark. The Julia code
generator will try to represent values by machine-native types. An integer
variable for example will often be lowered to a native integer. However, when
the variable hosts a complex object, or its type is dynamic, i.e., not known at
compile time to be a singular type, the value will be boxed, heap-allocated, and
garbage collected. On a GPU, support for garbage collected objects is hard to
implement: The current heap bookkeeping is inherently single-threaded, and lives
entirely in CPU memory. So in our framework, we completely depend on Julia to
lower data types to its native counterparts that won't be heap-allocated. If the
value cannot be represented natively, and hence would be boxed, compilation is
aborted. Although this might seem like a big restriction, kernel code is often
reasonably simple, only performing data transformations or other calculations.
Additionally, the Julia developers are continuously working on improving the
type lowering mechanics, as using native types rather than garbage-collected
boxes allows LLVM to reason about these types and therefore to optimize the code
better.

\footnotetext{The new Unified Memory solves this, but moves the synchronization
burden to the API, yielding lower performance than manual
management~\cite{accelware2014unified}.}

This dependence on Julia's type lowering does not prevent developers from
writing kernel code in a dynamic fashion: Variable types only need to be known
at compile time to avoid object boxing. Source code can still use dynamically
typed variables, and rely on type inference to derive the necessary type
information. If at some point a dynamically typed function is invoked with
previously unseen types of arguments, this triggers a new compilation. Type
inference is performed again, and the resulting machine code is saved for reuse
in a method cache.

Out of necessity, we implemented the actual \ac{ptx} code generation support in
the C++ core of the Julia code generator. This facilitates interacting with the
LLVM APIs and Julia internals, which are both written in C++.\footnotemark
Additionally, this enables reuse of parts of the CPU code generator.

\footnotetext{A non-standardized ABI and compiler differences hamper interaction
with C++ code. In the future, the
\href{https://github.com/Keno/Cxx.jl}{\texttt{Cxx.jl}} package could be used to
let an actual C++ compiler generate glue code. Early experiments have yielded
interesting results, but the package was not ready for practical use yet.}

Besides generating the \ac{ptx} code, executing Julia code on GPUs also requires
the hardware to be configured, the code to be uploaded, etc. We decided against
implementing this in the C++ code generator, instead extending a Julia package
as detailed in \Cref{api-wrapper,automation}. This has a twofold advantage:
Maintainability is greatly improved, and the added functionality is only enabled
when explicitly importing the package.

The Julia code in this package needs to invoke the code generator to compile
certain kernel functions to \ac{ptx} code. Regular calling mechanics do not
apply here, since calling a kernel requires invoking the CUDA APIs. We therefore
expose this functionality through \emph{compiler hooks}. These are intrinsic
functions predefined by the Julia compiler that enable Julia host code to call
into the code generator. These hooks are not only used to invoke the \ac{ptx}
code generator, but also to initialize and configure the LLVM \ac{ptx} target,
extract the compiled kernel's symbol name, and other auxiliary tasks.

\subsection{\texttt{@target} macro}
\label{julia-kernels:target}

As the compilation of GPU functions differs significantly from ordinary CPU
functions, a developer needs to be able to mark functions as such. We opted for
a lightweight language extension: a \texttt{@target} macro to preface a function
definition, as shown in \Cref{lst:julia_new} on \cref{lst:julia_new:target}.
This macro puts the specified target name as a meta attribute in the \ac{ast} of
the function. The compiler reads this attribute, and uses it to activate the
conditional, \ac{ptx}-specific compilation flow.

\section{CUDA API Wrapper}
\label{api-wrapper}

To launch kernels and perform all related tasks, we need access to a wide range
of CUDA API functions. Targeting the driver API rather than the higher-level
run-time API, we started from the existing \texttt{CUDA.jl}
wrapper~\cite{julia2015cuda}. This package provides a Julia interface for most
API calls, which we extended for our needs. For example, we added support for
shared memory and global variables in the form of idiomatic Julia constructs
rather than plain API call wrappers.

Additionally, we improved the package's compatibility by adding support for
different versions of the CUDA APIs. The package now also supports the
GPU~Ocelot emulator~\cite{farooqui2011gpuocelot}, which we extended in order to
cover all required API calls. Developers can now use the Julia GPU support
without having any physical NVIDIA hardware; This is interesting both for ease
of development, e.g., on a laptop or a continuous integration server, as well as
compatibility, e.g., to support GPUs of different vendors.

As part of the API, we defined intrinsic functions for use within GPU kernels.
Some of these intrinsics translate to \ac{ptx}-specific functionality exposed by
means of LLVM intrinsics, e.g., to access the size of and position within the
currently active block and grid. Some examples are shown on
\cref{lst:julia_new:intrinsics} of \Cref{lst:julia_new}. These are essential to
GPU kernels, which rely on that very position information to differentiate
parallelized computations between individual threads.

Other intrinsics map onto parts of the CUDA device library, \texttt{libdevice},
which contains serialized LLVM \ac{ir} code implementing common functionality
for each generation of GPU~\cite{nvidia2015cudadoc}. In some cases, this
functionality replaces built-in Julia counterparts unfit for GPU execution. For
example, the Julia standard library implements most basic trigonometric
functions by invoking the \texttt{openlibm} math library.  This library is not
available for execution on the GPU, however, and it would prove extremely costly
to hand control back to the CPU for every invocation. By using
\texttt{libdevice} instead, the computation can be performed on the GPU itself.

We implemented most of these intrinsic functions using the experimental
\texttt{llvmcall} functionality in Julia. This compiler feature allows a
developer to embed raw LLVM \ac{ir} code within Julia source code, similar to C
developers using inline assembly in their C code.

Sometimes, the functional behavior of a \texttt{libdevice} function or LLVM
\ac{ptx} intrinsic did not conceptually match what a Julia developer would
expect. For example, the aforementioned position information within a block or
grid is indexed starting with index 0, while Julia arrays are indexed starting
from 1. The intrinsics we expose for use within Julia code correct this
behavior, making them adhere to Julia conventions. This minimizes the
differences between host and device code, and allows the use of idiomatic
1-indexed array expressions in device code, e.g., based on the thread's position
within a block or grid.

\section{High-Level Automation}
\label{automation}

With the \ac{ptx} compiler support and the CUDA API wrapper package, it is now
possible to target NVIDIA GPUs and other accelerators entirely from within
Julia. No external compiler is required, and several commonly-used APIs are
wrapped using language-native constructs. An application relying on the already
described functionality would look like the code in \Cref{lst:julia_original},
but with the kernel from \Cref{lst:cuda} written in Julia as well. Although this
improves usability, knowledge of the CUDA API is still required, and certain
inherent restrictions make it difficult for layman developers to successfully
execute Julia code on a GPU.

\subsection{\texttt{@cuda} macro}
\label{automation:macro}

\begin{figure*}
  \centering
  \includegraphics{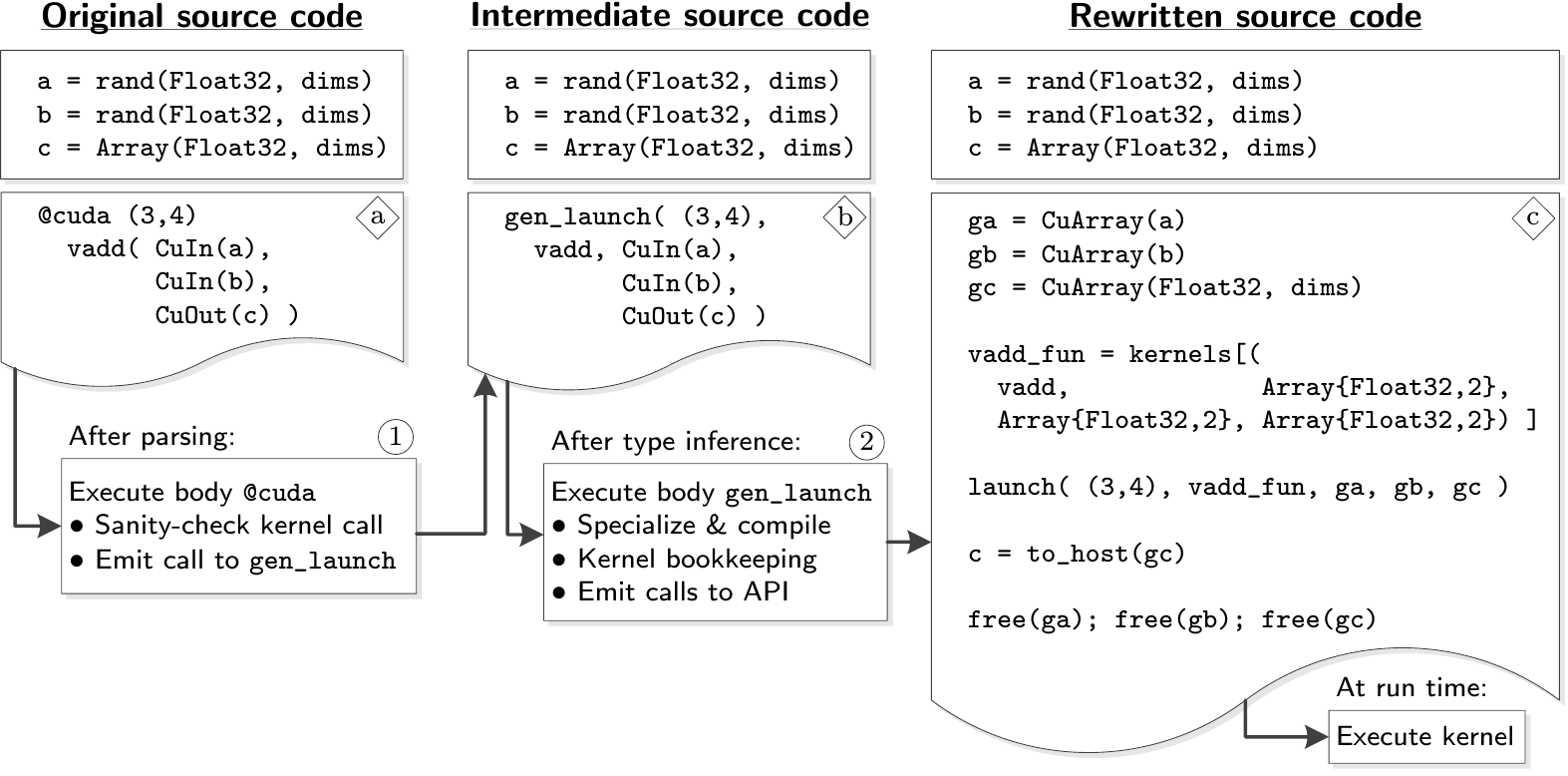}
  \caption{Example of code expansion resulting in a kernel launch.}
  \label{fig:expansion}
\end{figure*}

To hide the remaining complexity as much as possible, we developed a
\texttt{@cuda} macro to prefix kernel calls. The macro uses Julia's
powerful metaprogramming abilities discussed in
Section~\ref{background:julia:metaprogramming}, and mimics the well-known triple
 angle bracket calling syntax of CUDA~C, which looks as follows:
{ \small
\begin{verbatim}
   vadd<<<3,4>>>(a, b, c)
\end{verbatim}
}
Our \texttt{@cuda} macro looks similar, but refrains from using triple angle
brackets which the Julia parser does not accept:
{ \small
\begin{verbatim}
   @cuda (3,4) vadd(a, b, c)
\end{verbatim}
}

Contrary to the CUDA~C calling syntax that narrowly maps onto the kernel launch
API call, our macro also takes care of specializing and compiling the function,
configuring the hardware and launching the kernel. This is realized with minimal
developer interactions: using a version of the Julia compiler with \ac{ptx}
support and importing the \texttt{CUDA.jl} package is sufficient to enable this
functionality.

Behind the scenes, the \texttt{@cuda} macro results in two phases of processing
and code generation, denoted by \textcircle{1} and \textcircle{2} in
\Cref{fig:expansion}. First, when the Julia parser expands the macro invocation
from \textdiamond{a} (passing along expression trees encoding the dimension
information and the function call) the syntax of the kernel invocation is
checked, and code as in \textdiamond{b} is emitted to call the
\texttt{gen\_launch} generated function.

The generated function implements phase~\textcircle{2}, performing most of the
actual work. Much like the first-phase macro, the generated function returns
code in the form of an \ac{ast}, but its body is only executed after type
inference, when the arguments' types are known. This type information is
crucial. It allows the compiler to emit the exact amount of low-level glue code
required for executing the given Julia function on GPU hardware. For our running
example, this results in the code marked as \textdiamond{c} in
\Cref{fig:expansion}.

Using these metaprogramming mechanics rather than ordinary functions has several
advantages. First, it allows transparent use of the familiar calling syntax as
can be seen in \textdiamond{a}. The generated function will replace this with
the actually required API interactions from \textdiamond{c}. More importantly,
it avoids any run-time overhead: Each invocation of the \texttt{@cuda} macro and
ensuing call to \texttt{gen\_launch} are only executed once for every set of
argument types. The resulting code is saved in a method cache, and reused in
each subsequent invocation. The macro nor the generated function end up in the
final machine code, only the specialized glue code and the necessary CUDA API
calls are found there. Combined with Julia's \ac{jit} compiler, this should
allow for run-time performance comparable to that of statically-typed,
ahead-of-time compiled CUDA~C.

\subsection{Kernel specialization and compilation}
\label{automation:compilation}

Specializing the GPU kernel consists of generating a \emph{type-lowered}
version, where expression types have been deduced by means of type inference.
Generating this function is done by calling into the Julia compiler, using
pre-existing hooks. After this, all variables should be statically typed. If
this is not the case, the \ac{ptx} code generator will abort, because it would
result in unsupported, heap-allocated boxes.

Given a function's type-lowered \ac{ast}, the \ac{ptx} code generator is invoked
using the compiler hooks from \Cref{julia-kernels:codegen}. This involves
initializing the LLVM \ac{ptx} back-end (setting the triple, selecting an
architecture, \ldots), and generating \ac{ptx} code for all functions related to
the kernel invocation: the kernel function itself, as well as any callee that
was not inlined. This functionality is part of the \texttt{gen\_launch}
generated function, which is executed during phase~\textcircle{2}. Consequently,
even when the resulting kernel code will be executed multiple times, this
functionality does not need to be re-executed.

Finally, the generated function emits the necessary CUDA API calls for creating
a module, loading the \ac{ptx} code, acquiring a function pointer and launching
the kernel. Some of these API calls, such as creating and loading the module,
are evaluated at compile time, with the results cached for future invocations.
This further reduces the run-time overhead, but only applies to a small number
of API calls. For our example, the remaining calls, wrapped with constructs from
\Cref{api-wrapper}, are shown in code fragment~\textdiamond{c}.

\subsection{Argument conversion and management}
\label{automation:arguments}

In addition to compiling and managing the kernel's code, the generated function
also emits code that takes care of synchronizing the kernel's arguments between
the CPU and GPU, as can be seen in fragment~\textdiamond{c}. This involves
allocating and deallocating memory, converting values if necessary, and copying
data to and from device memory.

By default, all arguments will be uploaded to the device before execution, and
downloaded back to host memory after completion. Often, however, some arguments
will only be read from, while others serve as a container for return values.
Although this could be deduced through dataflow analysis of the kernel's
function body, we currently require the developer to specify this behavior using
wrapper classes. By optionally wrapping arguments with a \texttt{CuIn},
\texttt{CuOut} or \texttt{CuInOut} constructor, as shown in
fragment~\textdiamond{a} and on \cref{lst:julia_new:cuda} of
\Cref{lst:julia_new}, the developer can force the compiler to generate only the
absolutely necessary memory transfers.

\section{Evaluation}
\label{evaluation}

\subsection{Benchmarks}

The main contribution of this paper is a framework that compiles high-level
Julia code to \ac{ptx} instructions, including a library that wraps and
automates the CUDA API using abstractions that should not result in run-time
overhead. Testing our approach using conventional accelerator-oriented
benchmarks suites such as Parboil~\cite{stratton2012parboil} or
Rodinia~\cite{che2009rodinia} is not ideal: it would mainly stress the quality
of the underlying code generator, in our case the LLVM \ac{ptx} back-end, and it
would require porting each benchmark to Julia.

Instead, we evaluate our work by porting a real-life, scientific application
that uses GPUs in a multifaceted manner. We selected the trace transform, an
image processing algorithm that extracts image descriptors by projecting along
straight lines of an image in multiple orientations~\cite{kadyrov2001trace}. The
trace transform benchmark features a lot of parallelism, including
coarse-grained parallelism for processing different orientations concurrently
and more fine-grained parallelism to break up individual projections into
semi-independent computations~\cite{besard2015tracetransform}. Moreover, this
benchmark has been used before to compare programmer productivity and
performance for a wide range of programming languages, including MATLAB, MEX,
Octave, Scilab, C++, OpenMP, CUDA, and pure Julia (i.e., for CPUs only).

That existing work provides a well-optimized CUDA implementation and a baseline
Julia version. The code for these implementations is available on GitHub under
an open-source license. The CUDA version splits the algorithm in five or more
separate kernels, depending on the invocation mode. Some of these kernels are
simple and independent, while others feature complex computations and use shared
memory for inter-thread communication. We re-used the pure Julia version as the
starting point for our Julia GPU implementation, and compare it against the C++
with CUDA version to assess a potential loss in performance.

\subsection{Methodology}

All benchmarks are run on a dedicated 64-bit Debian Linux 3.16 system containing
an Intel i7-3770K processor (3.5 GHz, 4 cores, 8 threads) with 16 GB DDR3 main
memory, and an NVIDIA GeForce GTX Titan GPU. We compiled C++ sources with GCC
4.9.2, and used CUDA toolkit version 6.0.37 paired with NVIDIA driver version
340.65. All statically compiled C++ and CUDA code is optimized using the
compilers' \texttt{-O3} flags, disabling any debug functionality with
\texttt{-DNDEBUG}. In the case of CUDA code, we specifically compiled for our
GPU's architecture, \texttt{sm\_35}. Julia code is executed using our modified
version of the compiler, based on the 0.4 development tree from early 2015.

Benchmarks are run multiple times, discarding initial warm-up iterations. We
estimate the precision of the measurements by means of the relative uncertainty,
calculated on the basis of the standard deviation and mean of a log-normal
distribution~\cite{mashey2004benchmarks, ciemiewicz2001mean}. It is generally
accepted that relative uncertainties below 2\% are characteristic of careful
measurements~\cite{taylor1997error}. The measurements reported in this paper are
the means of a fitted log-normal distribution, with the maximum uncertainty
listed in the accompanying captions.

We compare the performance of 5 implementations, as listed in
\Cref{tbl:init,tbl:sloc}. The first three implementations are part of existing
work~\citep{besard2015tracetransform}. The fourth implementation re-uses the
(statically compiled) CUDA~C kernels from the second C++ version, in combination
with the pre-existing CUDA API wrapper to manage driver interactions. The last
version uses our framework to its full extent.

\subsection{Application performance}

The most important performance metric is the steady-state execution time. To
measure this, we modify the applications and place the main algorithm invocation
in a loop that measures the time it takes to execute each iteration.

\begin{figure}[!t]
  \centering
  \include{images}
  \label{fig:images}
\end{figure}

The results of these measurements are visualized in \Cref{fig:images}. It shows
how the CPU-based implementations scale linearly. By contrast, implementations
using the GPU scale superlinearly for small input sizes. This is due to the
large constant overhead of configuring the GPU and launching the kernels,
irrespective of the input size~\cite{besard2015tracetransform}.

The difference between the CPU-only C++ and Julia implementations scales
together with the image size, and is caused by deficiencies in the Julia code
generator, more specifically unnecessary checks on integer conversions and array
bounds~\cite{besard2015tracetransform}. These checks report failure by throwing
an exception, which is currently not supported by our \ac{ptx} code generator.
For the time being, we have disabled these checks altogether, therefore avoiding
this loss of performance when compiling Julia kernels to \ac{ptx} code.

In the case of the Julia implementation reusing statically compiled CUDA~C
kernels, the slowdown compared to C++ with CUDA is 13\% for small images, but
this drops to 2\% for more realistically sized inputs. The constant part of this
overhead, contributing more significantly at lower images sizes, can be
attributed to lower generated code quality of the inevitable Julia host code
between kernel launches. The remaining overhead at larger image sizes comes from
argument conversions, copying and converting Julia datatypes before they are
uploaded to GPU device memory. As the amount of data to copy scales in terms of
the input image size, this overhead does not disappear for larger images.

Finally, the Julia implementation targeting GPUs directly, i.e., with our
framework compiling Julia code to \ac{ptx} code on the fly, only runs slightly
slower, resulting in a 1.5\% extra overhead for all image sizes (compared to
using statically compiled CUDA~C kernels in Julia). Detailed measurements using
NVIDIA's profiling utility shows that this overhead is entirely caused by
differences in the code quality between code generated by the LLVM back-end we
use in our framework and code generated statically by the CUDA toolkit's
\texttt{nvcc} compiler. This overhead is hence not introduced by our approach
and implementation, and would vanish if LLVM simply became as good as
\texttt{nvcc} for compiling code to \ac{ptx}. We can therefore conclude that the
code generated by our automation framework from \Cref{automation} is on par with
manual API interactions.

\subsection{Program initialization}

\begin{table}[!t]
  \centering
  \small
  \nocaptionrule \caption{Static build and run-time initialization times for
  different implementations (relative uncertainty: 1.10\%).}
  \begin{tabular}{l c c}
    \hline
                              & \textbf{Build (s)}  & \textbf{Init (s)}       \\
    \hline
    C++ (CPU)                 & $13.36$             & $0.002$                 \\
    C++ (CPU)   + CUDA (GPU)  & $18.71$             & $0.012 $                \\
    \hdashline
    Julia (CPU)               &                     & $14.43$                 \\
    Julia (CPU) + CUDA (GPU)  & $1.26 $             & $14.66$                 \\
    Julia (CPU  + GPU)        &                     & $15.85$                 \\
    \hline
  \end{tabular}
  \label{tbl:init}
\end{table}

Although steady-state performance will be of most interest to users, it is
interesting to consider the build and initialization times required to start an
application from its sources. These measurements are shown in \Cref{tbl:init}.
Where applicable, we measured the time required for statically building an
application and any required objects such as GPU kernels, using a single thread
for the sake of comparison. In order to determine the initialization time, we
inserted code aborting executing after a single warm-up iteration. Subtracting
the known steady-state iteration time, this yields the time to initialize and
warm-up the application.

As expected, static compilation takes a significant amount of time, but avoids
run-time initialization overhead. Building the C++ with CUDA implementation
takes longer, as kernels and some auxiliary support classes need to be build in
addition to the main application's code. In the case of Julia with CUDA, these
auxiliary classes are part of the dynamically compiled CUDA API wrapper.
Therefore the static build time is shorter than for the C++ with CUDA version.

Compared to statically-compiled C++, dynamically compiled Julia code involves
a long initialization time. However, most of this overhead is due to loading
auxiliary packages, such as \texttt{Images.jl} for image support and
\texttt{ArgParse.jl} for parsing command-line arguments. New versions of Julia
are bound to reduce this overhead, as package precompilation is actively being
worked on. Compiling kernels at parse time adds only a relatively small
initialization overhead of around 8\%, increasing the initialization time from
14.66 to 15.85s. This is significantly shorter than the 5.35s added to the C++
build time when statically compiling the CUDA~C kernels. Furthermore, the
initialization time can be reduced by caching compiled kernels over multiple
executions.

\subsection{Programmer productivity}

To assess the programmer productivity, we count the number of source code lines
required to port a trace transform implementation to a GPU. As listed in
\Cref{tbl:sloc}, targeting a GPU from C++ by using the CUDA system headers and
writing kernels in CUDA~C requires a significant amount of work. The program
grows from 721 to 1184 lines of code, an increase of almost 65\%. Out of those
additions, only 163 lines concern the core algorithm, which grows from 69 lines
to 89 lines of C++ and 143 lines of CUDA~C. This illustrates how, in a
traditional C++ with CUDA programming environment, most of the porting effort
concerns auxiliary tasks not related to the actual computational part of the
application.

\begin{table}[!t]
  \centering
  \small
  \nocaptionrule \caption{Lines of code of entire application and core algorithm
  for different implementations.}
  \begin{tabular}{l c c c}
    \hline
                                & \textbf{Program}  & \multicolumn{2}{c}{%
                                                      \textbf{Core algorithm}}  \\
                                &                   & CPU   & GPU               \\
    \hline
    C++ (CPU)                   & $721$             & $69$  &                   \\
    C++ (CPU)   + CUDA (GPU)    & $1184$            & $89$  & $143$             \\
    \hdashline
    Julia (CPU)                 & $359$             & $49$  &                   \\
    Julia (CPU) + CUDA (GPU)    & $548$             & $80$  & $143$             \\
    Julia (CPU  + GPU)          & $449$             & $20$  & $108$             \\
    \hline
  \end{tabular}
  \label{tbl:sloc}
\end{table}

Using our version of the \texttt{CUDA.jl} wrapper to launch kernels from within
Julia, the implementation's code grows from 359 to 548 lines of code. This
addition of 189 lines is far less than what is required to launch the same
kernels from C++, as a result of the wrapper package taking care of auxiliary
tasks such as error checking and memory management.

If we replace the CUDA~C kernels with Julia code and use our framework to
compile and execute that code, we benefit from multiple advantages: the kernel
implementations are more concise, and most CUDA API interactions are automated.
This shortens the application by almost 100 lines compared to the Julia with
CUDA version. Most of the porting effort is now concentrated in the core
algorithm, which matters most in terms of application performance. On top of
that, all code is now written in the same programming language, further lowering
the barrier to programming GPUs.

\section{Related Work}
\label{related-work}

In recent times, many developments have added GPU support to general purpose,
high-level languages without depending on a lower-level, device specific
language such as CUDA or OpenCL. One popular approach is to host a DSL in the
general-purpose language, with properties that allow it to compile more easily
for GPUs. For example, Accelerate defines an embedded array language in
Haskell~\citep{chakravarty2011accelerate}, while Copperhead works with a
functional, data-parallel subset of Python~\citep{catanzaro2011copperhead}.
Parakeet uses a similar Python subset, with less emphasis on the functional
aspect~\citep{rubinsteyn2012parakeet}. Other research defines entirely new
languages, such as Lime~\citep{dubach2012lime} or
Chestnut~\citep{stromme2012chestnut}. In all these cases, the user needs to gain
explicit knowledge about this language, lowering his productivity.

An alternative to defining a new language is directive-based programming within
the host language. OpenACC extends C++~\citep{wienke2012openacc}, but only a
select number of compilers can compile it for GPUs.
jCudaMP~\citep{dotzler2010jcudamp} and
ClusterJaMP~\citep{veldema2011clusterjamp} feature OpenMP-like directives in
Java, subsequently compiled to low-level device code. Directive-based
programming is often portable over multiple devices, but this generality comes
at a cost, and complicates low-level optimization~\citep{hoshino2013cuda}.
hiCUDA avoids this by defining lower-level device-specific directives, which
require the programmer to be familiar with the CUDA data and execution
model~\citep{han2011hicuda}.

Our work allows writing kernels in the high-level source language, hiding the
underlying CUDA API from the developer. Rootbeer accomplishes a similar goal
within Java, but requires manual build-time actions to post-process and compile
kernel source code~\citep{pratt2012rootbeer}. Jacc features automatic run-time
compilation and extraction of implicit parallelism, but requires the programmer
to construct manually an execution task-graph using a relatively heavy-weight
API~\citep{clarkson2015jacc}. The new Rust language also gained lightweight GPU
support in a manner very similar to our work, featuring a comparable high-level
API wrapper on top of the existing OpenCL support~\citep{holk2013rustgpu}. Using
the wrapper is mandatory though, while our work attempts to automate most
interactions. In addition, all these projects build upon statically typed source
languages. Our work allows for dynamically typed kernels, automatically
specialized based on argument types. NumbaPro, being a Python compiler, does
support dynamically typed kernels~\citep{continuum2015numbapro}. It optionally
takes care of API interactions as well, but does not take into account how
arguments are used. NumbaPro also uses a custom Python compiler, which
significantly complicates the implementation and is currently not fully
compatible with the Python language specification.

\section{Conclusion and Future Work}

We presented a framework for high-level GPU programming in the Julia programming
language. It specializes and compiles Julia kernels to \ac{ptx} code using the
LLVM infrastructure. We extended the \texttt{CUDA.jl} API wrapper, adding
user-friendly abstractions and improving compatibility, even with non-CUDA
hardware. Finally, we support automating all API interactions, greatly improving
usability of GPUs.

We demonstrated the viability of our framework by porting a real-life image
processing algorithm, and comparing with pre-existing C++ and CUDA
implementations. The performance impact of writing GPU kernels in Julia is
negligible, while the required programmer effort is significantly reduced.
Boilerplate API interactions have disappeared, and users can focus on porting
time-consuming computations.

Our work makes it possible to write portable, high-performance GPU code with low
effort. It could be used to port high-level functionality from Julia's standard
library, making it possible to transparently use any available GPU. Extensions
to our framework may be required in order to improve compatibility with boxed
objects and automatically detect the use of arguments, but in its current state
it already proved usable for porting real-life applications.

\acks

This work is supported by the Institute for the Promotion of Innovation by
Science and Technology in Flanders (IWT Vlaanderen), and by Ghent University
through the Concerted Research Action on distributed smart cameras.

%%%%%%%%%%%%%%%%%%%%%%%%%%%%%%%%%%%%%%%%%%%%%%%%%%%%%%%%%%%%%%%%%%%%%%%%%%%%%%%%
% Bibliography
%

% We recommend abbrvnat bibliography style.

\bibliographystyle{abbrvnat}

% TODO: the bibliography (`main.bbl') should be embedded for final submission.
\bibliography{main}

\begin{thebibliography}{29}
\providecommand{\natexlab}[1]{#1}
\providecommand{\url}[1]{\texttt{#1}}
\expandafter\ifx\csname urlstyle\endcsname\relax
  \providecommand{\doi}[1]{doi: #1}\else
  \providecommand{\doi}{doi: \begingroup \urlstyle{rm}\Url}\fi

\bibitem[Acceleware(2014)]{accelware2014unified}
Acceleware.
\newblock {CUDA} 6.0 {Unified Memory} performance, 2014.

\bibitem[Besard et~al.(2015)Besard, De~Sutter, Fr{\'\i}as-Vel{\'a}zquez, and
  Philips]{besard2015tracetransform}
T.~Besard, B.~De~Sutter, A.~Fr{\'\i}as-Vel{\'a}zquez, and W.~Philips.
\newblock Case study of multiple trace transform implementations.
\newblock \emph{Int. Journal of High Performance Computing Applications}, 2015.

\bibitem[Bezanson et~al.(2014)Bezanson, Edelman, Karpinski, and
  Shah]{bezanson2014julia}
J.~Bezanson, A.~Edelman, S.~Karpinski, and V.~B. Shah.
\newblock Julia: A fresh approach to numerical computing.
\newblock \emph{arXiv preprint arXiv:1411.1607}, 2014.

\bibitem[Catanzaro et~al.(2011)]{catanzaro2011copperhead}
B.~Catanzaro et~al.
\newblock Copperhead: Compiling an embedded data parallel language.
\newblock \emph{ACM SIGPLAN Notices}, 46\penalty0 (8):\penalty0 47--56, 2011.

\bibitem[Chakravarty et~al.(2011)]{chakravarty2011accelerate}
M.~M. Chakravarty et~al.
\newblock Accelerating {Haskell} array codes with multicore {GPUs}.
\newblock In \emph{Workshop on Declarative Aspects of Multicore Programming},
  pages 3--14, 2011.

\bibitem[Che et~al.(2009)]{che2009rodinia}
S.~Che et~al.
\newblock Rodinia: A benchmark suite for heterogeneous computing.
\newblock In \emph{Int. Symp on Workload Characterization}, pages 44--54. IEEE,
  2009.

\bibitem[Ciemiewicz(2001)]{ciemiewicz2001mean}
D.~M. Ciemiewicz.
\newblock What do you `mean': Revisiting statistics for web response time
  measurements.
\newblock In \emph{Computer Measurement Group Conf.}, pages 385--396, 2001.

\bibitem[Clarkson et~al.(2015)Clarkson, Kotselidis, Brown, and
  Luj{\'a}n]{clarkson2015jacc}
J.~Clarkson, C.~Kotselidis, G.~Brown, and M.~Luj{\'a}n.
\newblock Boosting {Java} performance using {GPGPUs}.
\newblock \emph{arXiv preprint arXiv:1508.06791}, 2015.

\bibitem[{Continuum Analytics}(2015)]{continuum2015numbapro}
{Continuum Analytics}.
\newblock {NumbaPro} {Python} compiler, 2015.

\bibitem[Dotzler et~al.(2010)Dotzler, Veldema, and Klemm]{dotzler2010jcudamp}
G.~Dotzler, R.~Veldema, and M.~Klemm.
\newblock {jCudaMP}: {OpenMP}/{Java} on {CUDA}.
\newblock In \emph{Int. Workshop on Multicore Software Engineering}, pages
  10--17, 2010.

\bibitem[Dubach et~al.(2012)]{dubach2012lime}
C.~Dubach et~al.
\newblock Compiling a high-level language for {GPUs}.
\newblock \emph{ACM SIGPLAN Notices}, 47\penalty0 (6):\penalty0 1--12, 2012.

\bibitem[Farooqui et~al.(2011)]{farooqui2011gpuocelot}
N.~Farooqui et~al.
\newblock A framework for dynamically instrumenting {GPU} compute applications
  within {GPU Ocelot}.
\newblock In \emph{Workshop on General Purpose Processing on Graphics
  Processing Units}, pages 1--9. ACM, 2011.

\bibitem[Han et~al.(2011)]{han2011hicuda}
T.~D. Han et~al.
\newblock {hiCUDA}: High-level {GPGPU} programming.
\newblock \emph{IEEE Trans. on Parallel and Distributed Systems}, 22\penalty0
  (1), 2011.

\bibitem[Holk et~al.(2013)]{holk2013rustgpu}
E.~Holk et~al.
\newblock {GPU} programming in {Rust}: Implementing high-level abstractions in
  a systems-level language.
\newblock In \emph{Parallel and Distributed Processing Symp. Workshops \& PhD
  Forum}, 2013.

\bibitem[Holy et~al.(2015)]{julia2015cudart}
T.~Holy et~al.
\newblock Julia wrapper for {CUDA} runtime {API}.
\newblock \url{https://github.com/JuliaGPU/CUDArt.jl}, 2015.

\bibitem[Hoshino et~al.(2013)]{hoshino2013cuda}
T.~Hoshino et~al.
\newblock {CUDA} vs {OpenACC}: Performance case studies with kernel benchmarks
  and a memory-bound {CFD} application.
\newblock In \emph{Int. Symp. on Cluster, Cloud and Grid Computing}, pages
  136--143, 2013.

\bibitem[{Julia developers}(2015)]{julia2015doc}
{Julia developers}.
\newblock {Julia} documentation, 2015.

\bibitem[Kadyrov and Petrou(2001)]{kadyrov2001trace}
A.~Kadyrov and M.~Petrou.
\newblock The trace transform and its applications.
\newblock \emph{IEEE Trans. on Pattern Analysis and Machine Intelligence},
  23\penalty0 (8):\penalty0 811--828, 2001.

\bibitem[Lattner and Adve(2004)]{lattner2004llvm}
C.~Lattner and V.~Adve.
\newblock {LLVM}: A compilation framework for lifelong program analysis \&
  transformation.
\newblock In \emph{Int. Symp. on Code Generation and Optimization}, pages
  75--86, 2004.

\bibitem[Lin et~al.(2015)]{julia2015cuda}
D.~Lin et~al.
\newblock Julia programming interface for {CUDA}.
\newblock \url{https://github.com/JuliaGPU/CUDA.jl}, 2015.

\bibitem[Mashey(2004)]{mashey2004benchmarks}
J.~R. Mashey.
\newblock War of the benchmark means: time for a truce.
\newblock \emph{ACM SIGARCH Computer Architecture News}, 32\penalty0 (4), 2004.

\bibitem[{NVIDIA}(2015)]{nvidia2015cudadoc}
{NVIDIA}.
\newblock {CUDA Toolkit} documentation, 2015.

\bibitem[Pratt-Szeliga et~al.(2012)]{pratt2012rootbeer}
P.~C. Pratt-Szeliga et~al.
\newblock Rootbeer: Seamlessly using {GPUs} from {Java}.
\newblock In \emph{Int. Conf on High Performance Computing and Communication},
  pages 375--380, 2012.

\bibitem[Rubinsteyn et~al.(2012)]{rubinsteyn2012parakeet}
A.~Rubinsteyn et~al.
\newblock Parakeet: A just-in-time parallel accelerator for {Python}.
\newblock In \emph{USENIX Conf. on Hot Topics in Parallelism}, pages 14--14,
  2012.

\bibitem[Stratton et~al.(2012)]{stratton2012parboil}
J.~A. Stratton et~al.
\newblock Parboil: A revised benchmark suite for scientific and commercial
  throughput computing.
\newblock \emph{Center for Reliable and High-Performance Computing}, 2012.

\bibitem[Stromme et~al.(2012)]{stromme2012chestnut}
A.~Stromme et~al.
\newblock Chestnut: A {GPU} programming language for non-experts.
\newblock In \emph{Int. Workshop on Programming Models and Applications for
  Multicores and Manycores}, 2012.

\bibitem[Taylor(1997)]{taylor1997error}
J.~Taylor.
\newblock \emph{An Introduction to error analysis: The study of uncertainties
  in physical measurements}.
\newblock University Science Books, 2nd edition, 1997.

\bibitem[Veldema et~al.(2011)Veldema, Blass, and
  Philippsen]{veldema2011clusterjamp}
R.~Veldema, T.~Blass, and M.~Philippsen.
\newblock Enabling multiple accelerator acceleration for {Java}/{OpenMP}.
\newblock In \emph{USENIX Workshop on Hot Topics in Parallelism (HotPar’11),
  Berkeley, CA}, pages 1--6, 2011.

\bibitem[Wienke et~al.(2012)]{wienke2012openacc}
S.~Wienke et~al.
\newblock {OpenACC}: First experiences with real-world applications.
\newblock In \emph{Proc of the Int. Conf. on Parallel Processing}, pages
  859--870, 2012.

\end{thebibliography}
%\begin{thebibliography}{10}
%\softraggedright
%\end{thebibliography}

\end{document}